\documentclass[prb,twocolumn,aps,amsmath,showpacs,amsfonts,amssymb,10pt]{revtex4-1}
\usepackage{graphicx}
\usepackage{times}
\usepackage[utf8]{inputenc}
\usepackage[english]{babel}
\usepackage{fancyhdr}
\usepackage{marvosym}
\usepackage{hyperref}
\usepackage{amsmath}
\usepackage{mathrsfs}
\usepackage{latexsym}
\usepackage{amssymb}
\usepackage{graphicx}
\usepackage[dvips,dvipsnames,usenames]{color}
\usepackage{float}

\graphicspath{{/Figuras/}
{Figures_Graphene_xtResonance_Approximation/}
{Figures_Graphene_ttResonance_Approximation/}
{Figures_TI/}
{Figures_Graphene/}}

\newcommand{\ket}[1]{\vert #1 \rangle}
\newcommand{\bra}[1]{\langle #1\vert}

\begin{document}

\title{Dirac fermion time-Floquet crystal: manipulating Dirac points}
\author{Pablo Rodriguez-Lopez, Joseph J. Betouras, and Sergey E. Savel'ev}
\affiliation{Department of Physics, Loughborough University, Loughborough LE11 3TU, UK}

\begin{abstract}
We demonstrate how to control the spectra and current flow of Dirac electrons in both a graphene sheet and a topological insulator by applying either two
linearly polarized laser fields with frequencies $\omega$ and $2\omega$ or a monochromatic (one-frequency) laser field together with a spatially periodic static potential
(graphene/TI superlattice). Using the Floquet theory and the resonance approximation, we show that a 
Dirac point in the electron spectrum can be split into several Dirac points whose relative location in momentum space can 
be efficiently manipulated by changing the characteristics of the laser fields.
In addition, the laser-field controlled Dirac fermion band structure -- Dirac fermion time-Floquet crystal -- allows the 
manipulation of the electron currents in graphene and topological insulators. Furthermore, the generation of dc currents of desirable intensity in a chosen direction
occurs when applying the bi-harmonic laser field which can provide a straightforward experimental test of the predicted phenomena.
\end{abstract}
\maketitle

\section{Introduction}

A huge surge of interest for both graphene (e.g. \cite{Novoselov05, review} and references therein), and three dimensional topological insulators (TIs) with two-dimensional topologically protected surface (e.g. \cite{TI1} and references therein),  
has been stimulated by many unusual and sometimes counterintuitive properties of these
materials. Indeed, in both graphene and the surface states of TIs, the effective Hamiltonians describing an evolution of wave functions of electron elementary excitations are linear in the momentum, resulting in 
pseudo-relativistic phenomena (e.g., the Klein tunnelling \cite{paradox,paradox-exp}, the unconventional Hall effect \cite{Hall} or the nonlinear magnetization \cite{nonlinear}). 

In contrast to true relativistic particles which are difficult to manipulate, 
pseudo-relativistic Dirac fermions e.g. in graphene, can be controlled
by static periodic electric and/or magnetic fields, known as 
graphene superlattices (see, e.g., \cite{superl,super5}). Such nano-structures can be experimentally implemented to control 
both spectrum and transport properties of Dirac electrons in graphene. Similar kind of structures have been proposed and made in TIs \cite{Burkov, Jin, Li, Goyal}.

Alternatively, one can control the electron band structure and electron current both in graphene and in a topological insulator  
by applying a time-dependent laser field (see, e.g., \cite{laser1,laser2} for graphene).  It has been shown  \cite{laser2} that a monochromatic laser field 
splits Dirac cone energy spectrum into mini-zones which can either touch each other in several Dirac points or be separated by gaps depending
on electromagnetic field polarization. Analogous techniques can be applied for TIs.  
The fact that laser controlled graphene/TIs electron band manipulation is quite promising for applications \cite{laser2} as well as the search for new physics, created much recent activity \cite{Kitagawa, Lindner, Inoue, Yazyev}, while very recently the Floquet-Bloch states were observed on the surface of a TI \cite{Science} and a photonic Floquet crystal has been also proposed \cite{TI-Floquet}.

Even more intriguing Dirac fermion dynamics can occur when a laser field is applied to graphene/TI superlattices resulting in acquiring
an effective mass by the fermions \cite{laser2,efetov}. This situation has not been well studied yet due to a complex space-time dynamics described by the partial differential equation which cannot be reduced to a set of ordinary differential equations as for the case when either only laser field or only 1D periodic 
potential is applied. Recently, a new method for dealing with such a situation has been proposed \cite{ph_prl} and 
giant backscattering resonances for electrons with small incident angles with respect to  
a 1D potential barrier has been predicted. This makes any further research in the field of laser-driven
graphene/TI superlattices very timely. 
 
In the present study, we first show that the Hamiltonian for a TI in external electromagnetic fields can be transformed to the Dirac graphene Hamiltonian by multiplying the second component of the corresponding spinor by an imaginary unity $i$, if the component of the vector potential in the perpendicular direction $A_z$ is zero. This makes both problems for graphene and TIs in electromagnetic fields mathematically equivalent and allows to unify all the developed techniques
for the manipulation of electrons in both materials. 

Then we focus on new unexplored phenomena, studying the manipulation of the Dirac energy cone by varying the characteristics of the field of two linearly polarized monochromatic lasers, one with frequency $\omega$ and the other with double frequency (i.e., $2\omega$). We demonstrate, using the first order resonance approximation (FORA), that the Dirac point of the original energy cone can be split into several Dirac points whose location in momentum space and even their number can be readily controlled by the angle between the two oscillating fields as well as the amplitudes and time-phase shift of the electric fields of two lasers. 

Moreover, our approach allows to estimate the time evolution of a wave function of Dirac fermions, and, thus,
calculate the fermion current at each state with certain momentum (or wave number ${\bf k}$). This resulting current is similar for both graphene and TIs and can be controlled by the time phase shift or the relative angle of two laser fields, allowing even to generate a dc current in a desirable direction due to the effect of harmonic mixing. We also consider a graphene/TI superlattice driven by a monochromatic (one-frequency) laser field and show that the number and location of
the Dirac points can be controlled by the relative angle between the laser and the static electric fields as well as their spatial and time periods of oscillations.

Finally, we go beyond the first order resonance approximation, for a simple case, to illustrate the validity of this approximation in the conclusions we arrive at.  

\section{Bi-harmonic laser for pristine samples}

In the low-energy limit, the behavior of
the charge carriers in electric field in graphene is described \cite{review} by the standard two-dimensional (2D)
Dirac equation where we set the electron charge $e=-1$.
\begin{equation}
i\partial_{t}\psi = {\bf {\sigma}}\centerdot\left({\bf p} - {\bf A}(t)\right)\psi, \label{eq1}
\end{equation}
with the two-component wave function $\psi= (\psi_A,\psi_B)$ for electrons in the two triangular sublattices, ${\bf p}=(-i\partial/\partial x, -i\partial/\partial y)$ is the momentum operator, ${\bf A}$ is the vector potential and the vector ${\bf {\sigma}}$ represents the Pauli matrices $\sigma_x$ and $\sigma_y$ (hereafter, we set $\hbar=1, v_F=1$ where $v_F$ is the Fermi velocity). For the case of a topological insulator,
the equation for evolution of a wave function $\phi$ is the same if we replace the operator ${\bf \sigma}\centerdot\left({\bf p} - {\bf A}(t)\right)$ by 
$\hat{z}\centerdot{\bf \sigma}\times\left({\bf p} - {\bf A}(t)\right)$, where $\hat{z}$ is a unit vector pointed perpendicular to the 2D plane. However, both
equations for the Dirac-like fermions in both graphene and a TIs coincide, if we use the following substitution $(\psi_{A},\psi_{B})=(\phi_{A}, i\phi_{B})$ in the equation for TIs. 
Similar analysis shows that fermion currents for graphene and TIs completely coincide if these are calculated for the corresponding states (e.g. states with a certain momentum). 
The reason is that one simply rotates the electron spin by multiplying the bottom component of a spinor by $i$, but this procedure does not affect the charge degrees of freedom. Thus, the calculations of the spectrum and single-particle currents described below are applicable for both TIs and graphene. In the following,  we focus on the Eq. (\ref{eq1}) only.

As a driving field ${\bf A}=(A_x(t),A_y(t))$ in Eq. (\ref{eq1}), we consider the superposition of two linear polarized electric fields having frequency $\omega$ and $2\omega$; namely:
\begin{eqnarray}
A_x &=& A_{1}\cos(\omega t)+ A_{2}\cos(\theta)\cos(2\omega t+\alpha), \nonumber \\ 
A_y&=&A_{2}\sin(\theta)\cos(2\omega t + \alpha).
\label{bi-field}
\end{eqnarray}
where $\bf{A_1}$ is oriented along the $x$ axis, without loss of generality. The angle $\theta$ is the one between the two oscillating fields with amplitudes proportional to $A_1$ and $A_2$, while $\alpha$ is a phase shift of the two ac drives
at time $t=0$. Such a driving field can be generated by two monochromatic lasers with a wave length much longer than the graphene (or TI) sample size projection on the weve propagation direction (in worst scenario the sample size should be smaller than 
laser wave length if wave propagates along the sample plane or just the sample thickness (several atomic layers) if the laser field propagates accross the sample). The time evolution of a wave function with a certain momentum ${\bf k}$ can be described by $\psi=\psi_0(t)\exp(ik_xx+ik_yy)$ and the Dirac equation (\ref{eq1}) can be reduced to a set of two ordinary
differential equations for the two components $\psi_{0A}(t)$ and $\psi_{0B}(t)$ of the spinor $\psi_0(t)$. Since the coefficients of the ordinary differential equations are periodic functions in time due to periodicity of
${\bf A}$, we can use the Floquet theory searching for a solution in the form $\psi_0(t)=\exp(i\varepsilon t){\tilde\psi}_0(t)$ where ${\tilde \psi}_0(t)$
is the periodic function, which can be
expanded in Fourier series, and $\varepsilon$ is the quasi-energy. In the resonance approximation (see, e.g. \cite{laser2}) we keep only the first lowest harmonics in the  Fourier expansion which are directly linked by the bi-frequency laser field.
Substituting the following expression
\begin{eqnarray}
\nonumber
\psi_0 = e^{-i \varepsilon t}\Biggl[\left(\begin{array}{c}
\psi_{A}^{++}\\
\psi_{B}^{++}
\end{array}\right)e^{+i\frac{3\omega}{2}t} + \left(\begin{array}{c}
\psi_{A}^{+-}\\
\psi_{B}^{+-}
\end{array}\right)e^{+i\frac{\omega}{2}t} + \noindent \\ + \left(\begin{array}{c}
\psi_{A}^{-+}\\
\psi_{B}^{-+}
\end{array}\right)e^{-i\frac{\omega}{2}t} + \left(\begin{array}{c}
\psi_{A}^{--}\\
\psi_{B}^{--}
\end{array}\right)e^{-i\frac{3\omega}{2}t}\Biggr], \label{bi-firstorder}
\end{eqnarray}
into eq. (\ref{eq1}) and equating the amplitudes that multiply $e^{-3i \omega t/2}$, $e^{-i \omega t/2}$,  $e^{i \omega t/2}$ and  $e^{3i \omega t/2}$ separately while ignoring higher harmonics, we arrive at a simple
linear matrix equation 
\begin{equation}
\varepsilon \bar{\psi}=L\bar{\psi} \label{mat-eq}
\end{equation} 
with $L$ a matrix that has constant elements and 
$\bar{\psi}=(\psi_{A}^{++}, \psi_{A}^{+-}, \psi_{A}^{-+}, \psi_{A}^{--}, \psi_{B}^{++}, \psi_{B}^{+-}, \psi_{B}^{-+}, \psi_{B}^{--})$ consisting of the time-independent amplitudes in Fourier series. This eigenvalue-eigenvector problem
can be solved numerically for each set of $k_x$ and $k_y$ allowing to construct both the electron pseudo-spectrum $\varepsilon(k_x,k_y)$ (see Fig.1 for the two lowest energy zones) and an approximate wave function $\psi=\psi_0(t)\exp(ik_xx+ik_yy)$  
for the Dirac fermions driven by the laser fields (eq.\ref{bi-field}). Moreover, the calculated approximate wave functions can be used to estimate the Dirac fermion one-particle currents (Fig. 2) at states corresponding to different values of momentum. 

Numerically, the calculated spectrum contains the first eight sub-bands. In order to consider the other sub-bands corresponding to higher energies, the use of a higher order resonance approximation is needed, keeping higher order harmonics in the ${\tilde \psi}_0$-expansion. In this section we
focus on the first two sub-bands touching each other in several Dirac points. One main result is that the number and location (in momentum space) of these points are controlled by the bi-harmonic laser field and can be manipulated by any one of three methods, by changing: (i) the relative orientation $\theta$ 
of the electric fields of the first and second laser harmonics, (ii) the relative time shift $\alpha$ of these drives, (iii) the drive amplitudes $A_1$ and $A_2$. 
These Dirac points originate from the splitting of the initial Dirac point connecting the positive energy Dirac cone with the negative energy Dirac cone, if the laser field is switched off.    

\begin{figure}[htb]
\includegraphics[width=9cm]{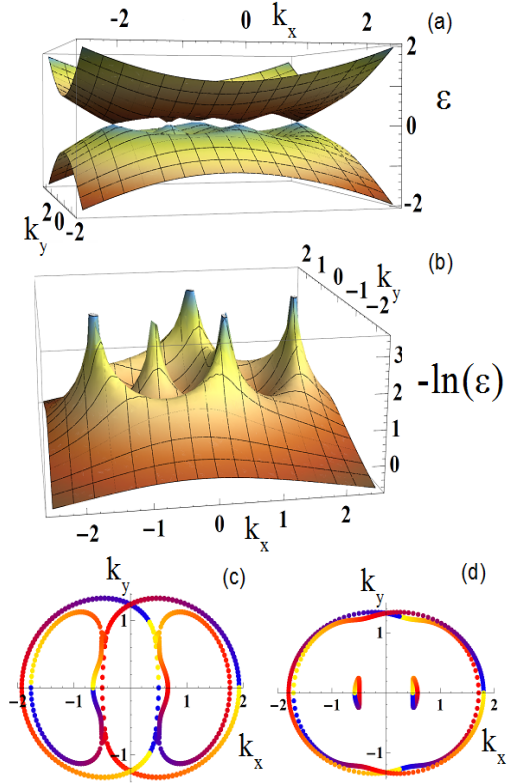}
\caption{(color online) (a) The two lowest zones of the quasi-energy spectrum $\varepsilon(k_x,k_y)$ when applying two-frequency laser field (\ref{bi-field}) with amplitudes $A_1=A_2=1$, time phase shift $\alpha=0$, and relative angle $\theta=\pi/10$. To highlight
five Dirac points where $\varepsilon=0$, we plot $F=-\ln[\varepsilon(k_x,k_y)]$ for the top energy zone in (b). Panel (c) shows the evolution of the locations of the Dirac points when changing the relative angle $\theta$ of the electric fields, i.e., the angle between ${\bf A}_1$ and ${\bf A}_2$;
different colours code different angles from blue for $\theta=0$ to yellow for $\theta=2\pi$, all other parameters are the same as in panel (a). Panel (d) is the same as panel (c) for the same set of parameters except for the weaker second field amplitude $A_2=0.5$.
}
\label{spectrum}
\end{figure}

Fig. 1(a) shows a representative 3D plot of the two lowest pseudo-energy zones $\varepsilon(k_x,k_y)$  touching in five Dirac points when the relative angle between two electric fields corresponding to the lasers with frequency $\omega$ and $2\omega$ is $\pi/10$ 
while these fields have the same amplitudes and zero time shift $\alpha$. In order to clearly see the Dirac points we replot (Fig. 1(b)) the same data for upper zone calculating $F=-\ln(\varepsilon)$ since the region of small values of $\varepsilon$ near the points $\varepsilon=0$ is highlighted in this representation.
By changing the relative orientation of the laser fields we can move the location of Dirac points  in the momentum space. Figure 1(c) shows the example of such a motion where different dots with different colours correspond to the positions $(k_{x}^D,k_{y}^D)$ of Dirac points (where $\varepsilon(k_{x}^D,k_{y}^D)=0$) 
for different orientations 
$\theta$ (video 1 can be provided upon request for an illustration).  Tracing the location of Dirac points we conclude that even the number of Dirac points can be changed by varying $\theta$ (e.g., from six Dirac points at $\theta=0$ to five Dirac points at $\theta$ near $\pi/10$ and then back to six Dirac points with further increasing $\theta$). 
The dynamics of the Dirac points with varying laser field parameters is quite rich and rather complicated. By decreasing the amplitude
of the second laser field, the dynamics of the Dirac points become less complex (Fig. 1 (d)) with a tendency towards a simple rotation of the Dirac points when the second laser is switched off, as already anticipated \cite{laser2}. 

\begin{figure}[htb]
\includegraphics[width=9cm]{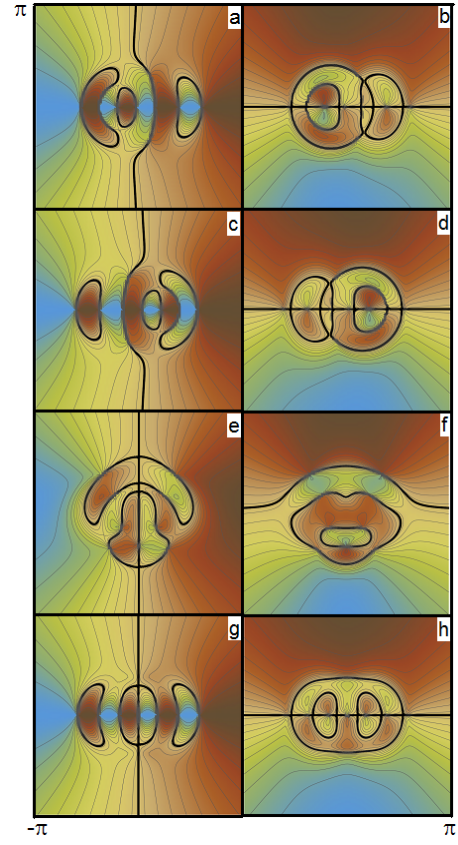}
\caption{(color online) Distribution of the single-particle currents $j_x(k_x,k_y)$ (the left column) and $j_y(k_x,k_y)$ (the right column) for the states with a fixed momentum $(k_x,k_y)$ when the harmonic field (\ref{bi-field}) with $A_1=A_2=1$ is applied. Four different cases are considered:
(a,b) $\theta=0$, $\alpha=0$; (c,d) $\theta=\pi$, $\alpha=0$; (e,f) $\theta=\pi/2$, $\alpha=0$; (g,h) $\theta=0$, $\alpha=\pi/2$. Depending on the relative angle $\theta$ and the time shift $\alpha$ of electric fields,
one can expect different symmetry of the current distributions and a possible generation of the DC electric current (see discussion in the text). The colors run from light blue when $j_i=1$ to dark red when $j_i=-1$ and the thick dark curves are the points $(k_x,k_y)$ with $j_i=0$.}
\label{currents}
\end{figure}

\begin{figure}[htb]
\includegraphics[width=9cm]{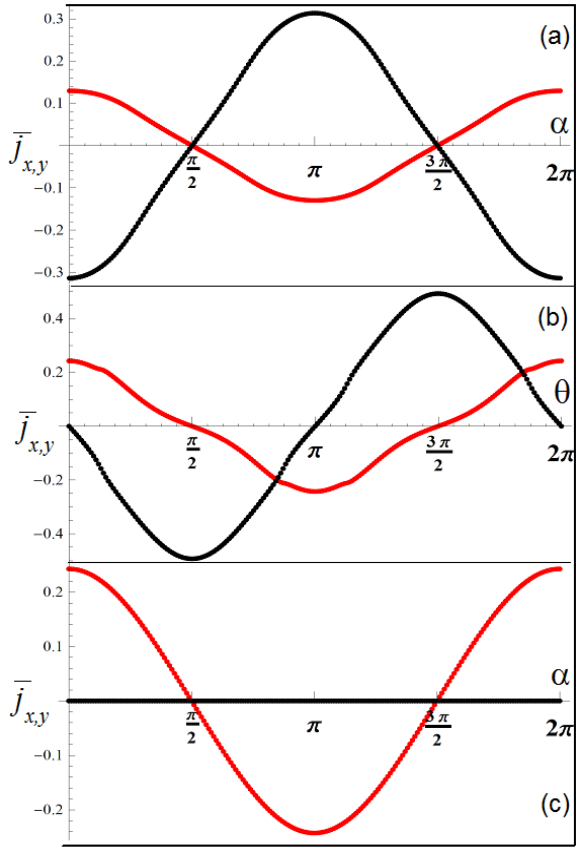}
\caption{(color online) The mean currents $\bar{j}_{x,y}$ for a certain magnitude of the momentum $k$, but averaged with respect to ${\bf k}$ orientations (see text). (a) Red and black dots correspond to $\bar{j}_x(\alpha)$ and $\bar{j}_y(\alpha)$ current components respectively, calculated for $A_1=A_2=1$, $k=0.25$, $\theta=\pi/4$ and $\alpha$
changing from 0 to $2\pi$; note that the nodes $\bar{j}_x(\alpha)=0$ and $\bar{j}_y(\alpha)=0$ coincide, giving $\theta=\pi/2+\pi n$ with integer $n$. (b)  Red and black dots show the $\bar{j}_x(\theta)$ and $\bar{j}_y(\theta)$ dependence, calculated for $A_1=A_2=1$, $k=0.25$, $\alpha=0$; the nodes $\bar{j}_x(\theta)=0$ 
and $\bar{j}_y(\theta)=0$ are shifted by $\pi/2$ giving the possibility to have a nonzero $\alpha$-dependence of one current component and zero value of the other component at the points $\theta=\pi n/2$ with integer $n$. Example of such a case is shown in (c) where $\bar{j}_x(\alpha)\ne 0$ and $\bar{j}_y(\alpha)=0$
for $A_1=A_2=1$, $k=0.25$, $\theta=0$. }
\label{DC-currents}
\end{figure}

\begin{figure}[htb]
\includegraphics[width=9cm]{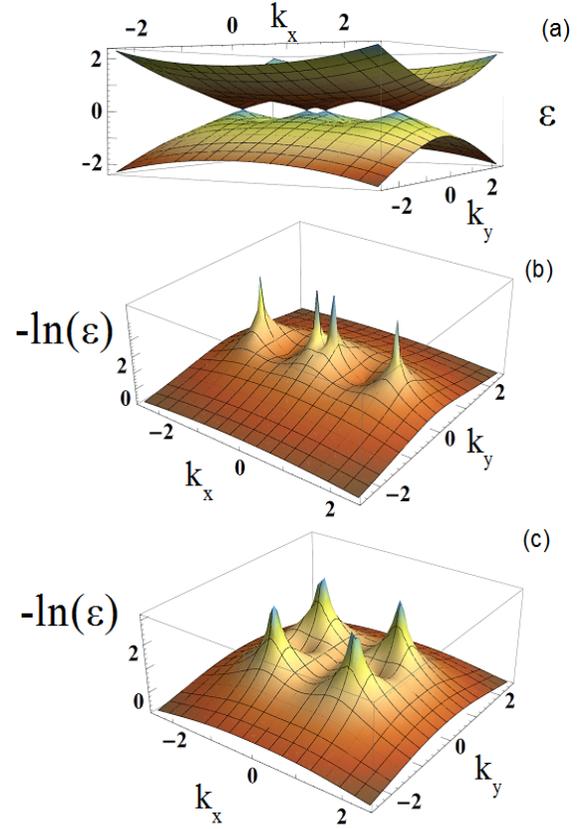}
\caption{(color online)  (a) The two lowest zones of the quasi-energy spectrum $\varepsilon(k_x,k_y)$ for graphene/TI superlattice (eq.(\ref{pot})) 
with $U_0=1$ and spatial period $L=2\pi/\mu=4\pi/3$, i.e., $\mu=1.5$,  when applying monochromatic laser field (\ref{one-field}) along 
the $x$-axis ($\theta=0$) with amplitudes $A_1=1$. To highlight
the four Dirac points where $\varepsilon=0$, we plot $F=-\ln[\varepsilon(k_x,k_y)]$ for the top 
energy zone in (b) for the same parameters as in (a) and in (c) for the same perameters as in (a) except the laser field orientation ($\theta=\pi/2$). }
\label{energy-super-currents}
\end{figure}

\begin{figure}[htb]
\includegraphics[width=9cm]{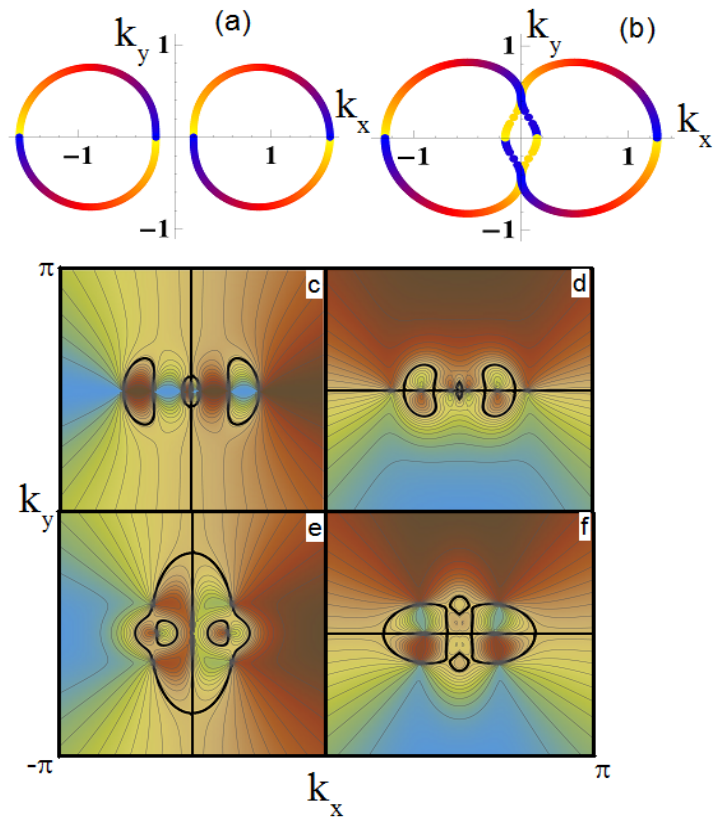}
\caption{(color online)  Panels (a,b) show the evolution of the locations of the Dirac points for graphene/TI superlattice when changing the relative angle $\theta$ of the monochromatic electric laser field (eq. \ref{one-field}) and electrostatic field $\nabla U$.
Other parameters are: $A_1=1$, $U_0=1$, $\mu=1.5$ for (a) and $\mu=0.5$ for (b); different colors code different angles from blue for $\theta=0$ to yellow for $\theta=2\pi$.
(c-f) Distribution of the single-particle currents $j_x(k_x,k_y)$ (the left column) and $j_y(k_x,k_y)$ (the right column) for the states with a fixed momentum $(k_x,k_y)$ when the monochromatic laser field (\ref{one-field}) is applied to a graphene/TI superlattice
for all parameters as in (a) while $\theta=0$ for (c,d) and $\theta=\pi/2$ for (e,f).}
\label{currents-super}
\end{figure}

Such rich dynamics of the Dirac points (especially when both amplitudes of laser fields, $A_1$ and $A_2$, are strong enough) has to significantly affect the transport properties of Dirac fermions at least at low energies. In order to prove this, we calculate the $x$ and $y$ components of the single-particle currents $j_{i} = \bra{\psi}\partial_{k_{i}}H\ket{\psi}$ at each state with certain value of the wave vector ${\bf k}=(k_{x}, k_{y})$ ($j_{i} = \psi^{*}\sigma_{i}\psi$ for graphene and $j_{i} = \phi^{*}\epsilon^{ji}\sigma_{j}\phi$ for TIs, where $\epsilon^{ji}$ is the Levi-Civita symbol in 2D). As demonstrated above, we obtain the same results for graphene and TIs, thus we focus on the case of graphene here. Several representative contour-plots of $j_x(k_x,k_y)$ (the left column) and $j_y(k_x,k_y)$ (the right column) is shown in Fig. 2. Figures 2(a,b) show the case when both electric fields are oriented along the $x$-axis destroying the reflection symmetry
with respect to the $y$-axis. Therefore, the condition $j(k_x,k_y)\ne - j_x(-k_x, k_y)$ would lead to a measurable DC current along the $x$-axis even within a proper many-body kinetic description (of course, an experimentally measured current should also depend on the distribution function which is beyond our
simple single-particle consideration, but the qualitative result that we describe here will be observed). In contrast, the symmetry $j_y(k_x,k_y)=-j_y(k_x,-k_y)$ should prevent any DC current along the $y$-axis. When rotating the 
$2\omega$ laser field by $\pi$ with respect to the $x$-axis, the $x$-current distribution also turns by 180$^\circ$ degrees indicating the change of sign of the $x$-axis DC current while the $y$-axis current is still zero, due to the survived relation $j_y(k_x,k_y)=-j_y(k_x,-k_y)$ (see Fig. 2(c,d)). 
It is interesting to note that the same current distributions can be obtained by either changing $A_2\rightarrow -A_2$ or $\alpha\rightarrow \alpha+\pi$. In other words, by adding $\pi$ either to $\theta$ (actual rotation of the electric field of the $2\omega$ laser) or to $\alpha$ (additional phase shift in 
the laser time dependence)  the changes of the currents are the same. In order to see the difference in the current distributions between the spatial rotation and the time shift of the laser fields, an extra $\pi/2$ can be added to either $\theta$  (Fig. 2(e,f)) or $\alpha$ (Fig. 2(g,h)). For a spatial rotation (Fig. 2(e,f)) the pattern in
the $k$ space is rotated by 90$^\circ$ degrees', resulting in the zero DC $x$-current due to the condition $j_x(k_x,k_y)=-j_x(-k_x,k_y)$ and a nonzero DC $y$-current since $j_y(k_x, k_y)\ne -j_y(k_x,-k_y)$. By shifting the time dependence by $\pi/2$ (Fig. 2(g,h)), while keeping $\theta=0$, both ($x-$ and $y-$) DC currents should 
become zero due to high symmetry of the obtained current distributions: $j_x(k_x,k_y)=-j_x(-k_x,k_y)$ and $j_y(k_x,k_y)=-j_y(k_x,-k_y)$.

All the above properties can be clearly seen, if we introduce a mean current for a certain magnitude $k$ of the momentum 
while averaged with respect to the momentum orientation: 
\begin{equation}
\bar{j}_{x,y}(k)=\int_0^{2 \pi} (d\gamma/2\pi)j_{x,y}(k\cos\gamma, k\sin \gamma)
\end{equation}
Such current represents the property of harmonic mixing of the electric current in graphene, driven by
two frequency laser field (Eq.(\ref{bi-field}), seeFig. 3). Interestingly,
the nodes $\bar{j}_{x}(\alpha)=0$ and $\bar{j}_{y}(\alpha)=0$ coincide resulting in zero of both the $x-$ and $y-$ components of the mean current at $\pi/2+\pi n$ with integer $n$ for any $\theta$. In contrast, the nodes of $\bar{j}_x(\theta)$ and $\bar{j}_y(\theta)$ are shifted by $\pi/2$, thus, resulting in 
$\bar{j}_x(\theta=\pi/2+\pi n)=0$ and $\bar{j}_y(\theta=\pi n)=0$ for any $\alpha$ and $k$. Note that such an unprecedented level of the DC current control by varying parameters of the two frequency drive is remarkable and provides further analogy with the classical \cite{hm1,hm2}, semi-classical \cite{hm3} 
and quantum \cite{hm-q} harmonic mixing. Recently, it was shown \cite{ph_prl} that a superposition of scalar potential barriers and time dependent laser fields can produce a resonant amplification of reflections of the Dirac fermions. This effect should also strongly amplify the harmonic mixing discussed here, allowing its experimental verification. A full kinetic description as well as consideration of damping and many body effects can partly hide the
property described here, of single-particle currents, which should be weighted with a proper non-equilibrium distribution function. Nevertheless, we believe that all the
symmetrical properties of the currents $\bar{j}_{x,y}$ that are described, should survive even in the proper kinetic description. 

\section{Monochromatic laser for superlattices}
\label{sec-3}
Here we consider a so-called graphene/TI superlattice where
a pristine  graphene/TI sample is modulated by a static periodic electrical field with a potential
\begin{equation}
U(x)=U_0\cos(\mu x) \label{pot}
\end{equation}
To manipulate the Dirac points and one particle currents in this superlattice we can apply a monochromatic laser field
\begin{equation}
A_x=A_1\cos(\theta)\cos(\omega t),\ \ A_y=A_1\sin(\theta)\cos(\omega t), \label{one-field}
\end{equation}
where the angle $\theta$ is between the laser field ${\bf A}(t)$ and the electrostatic field $\nabla U$ with the standard definition 
$\nabla=(\partial/\partial x, \partial/\partial y)$.

For the case of graphene superlattices in a laser field (see e.g., \cite{laser2}), the $y$-component of momentum conserves and the
solution of the Dirac equation 
\begin{equation}
i\partial_{t}\psi = [{\bf {\sigma}}\centerdot\left({\bf p} - {\bf A}(t)\right)+U(x)]\psi. \label{dirac-super}
\end{equation}
can be written in the form $\psi=\psi_0(x,t)\exp(ik_y y)$. As in previous section, we can introduce quasi-energy and quasi-momentum by using
Floquet-Bloch theory searching solutions $\psi_0$ in the form $\psi_0=\exp(i\varepsilon t+ik_x x)u(x,t)$ with $u$ being periodic in both time and space.
Therefore, we can again introduce a resonance approximation by expanding $u$ in the Fourier series with respect to both $t$ and $x$ and restricting the expression to several lowest harmonics directly linked via either the laser field or the electrostatic potential. In other words, we can search for $\psi_0(x,t)$
in the form:
\begin{eqnarray}
\psi_0 &=& e^{-i \varepsilon t+ik_x x}\times\nonumber \\
&\times&\Biggl[\left(\begin{array}{c}
\psi_{A}^{++}\\
\psi_{B}^{++}
\end{array}\right)e^{+i\frac{\omega}{2}t+i\frac{\mu}{2}x} + \left(\begin{array}{c}
\psi_{A}^{+-}\\
\psi_{B}^{+-}
\end{array}\right)e^{+i\frac{\omega}{2}t-i\frac{\mu}{2}x} + \noindent \nonumber\\ &+& \left(\begin{array}{c}
\psi_{A}^{-+}\\
\psi_{B}^{-+}
\end{array}\right)e^{-i\frac{\omega}{2}t+i\frac{\mu}{2}x} + \left(\begin{array}{c}
\psi_{A}^{--}\\
\psi_{B}^{--}
\end{array}\right)e^{-i\frac{\omega}{2}t-i\frac{\mu}{2}x}\Biggr] \label{mu-firstorder}
\end{eqnarray}
Substituting this expression in the Dirac equation (\ref{dirac-super}), the problem is reduced to the matrix equation (\ref{mat-eq}) with simply different
matrix elements comparing to the case studied in the previous section. Therefore, we can again solve the eigenvector-eigenvalue problem numerically and
derive an approximate quasi-energy spectrum as well as the corresponding approximate wave function $\psi$, which can be used to estimate the single-particle currents $j_x(k_x,k_y)$ and $j_y(k_x,k_y)$.  

Figure 4a shows the two lowest energy zone touching in the four Dirac points when the static periodic $\propto \nabla U$ and monochromatic laser
${\bf A}_1$ electric fields are both oriented along the $x-$axis. In this case (see also the highlighted representation of the Dirac point structure in
Fig. 4b where $F=-\ln(\varepsilon)$ is plotted), all the Dirac points are located on the $k_x$-axis. The rotation of the laser field by 90$^\circ$ results in a shift
of these Dirac points away from the $k_x$-axis (see Fig. 4c) [detailed dynamics can be seen 
upon request in video 2 which shows the motion and change of number of the Dirac points when gradually changing the angle $\theta$].

By changing the spatial period $L=2\pi/\mu$ of the superlattice potential $U$ compared with the corresponding scale $2\pi v_F/\omega$ [note 
that $v_F=1$ in the unit system we use here] of the monochromatic laser field, the different Dirac point dynamics can be observed (Fig. 5 a,b and 
the corresponding videos). For a short spatial period, $\mu>\omega$, of $U$, the two Dirac points moves along two well-separated almost-circle 
trajectories with no
intersections. Increasing the space period of $U$ results in touching trajectories for $\mu=\omega$ as well as more complicated, interconnected trajectories
(Fig. 5b) for $\mu<\omega$. For the last case, even number of the Dirac points can vary from 4 to 8 as seen in the available video. 

Regarding the one-particle current distribution $j_{x,y}(k_x,k_y)$ (see Fig. 5c-f), the obtained patterns are highly symmetric, $j_x(-k_x,k_y)=-j_x(k_x,k_y)$
and $j_y(k_x,-k_y)=-j_y(k_x,k_y)$, resulting in the zero dc-currents for any $\theta$. Nevertheless, the obtained patterns have a peculiar structure which can
affect some transport properties which are sensitive to the states with different fermion momentum. Also, the obtained patterns can be readily controlled by rotating the electric laser field with respect to the static field (compare Fig. 5(c,d) with Fig 5(e,f)).      

\section{Beyond the first order resonance approximation}

\subsection{Temporal second order resonance approximation for graphene in monochromatic laser field}

One of the simplest ways to assess the validity of the resonance approximation used above, is to calculate the Dirac points in a higher order resonance approximation keeping more harmonics in the expansion of ${\tilde \psi}_0(t)$ and observe the differences. Here we consider the simplest possible case,
with a single monochromatic laser field ${\bf A}=(0, A_1\cos\omega t)$ and no spatial modulations $U=0$. In the first order resonance approximation we search for $\psi=\psi_0(t)\exp(ik_xx+ik_yy)$  with
\begin{eqnarray}
\nonumber
\psi_0^{(1)} = e^{-i \varepsilon t}\Biggl[\left(\begin{array}{c}
\psi_{A}^{(1),+}\\
\psi_{B}^{(1),+}
\end{array}\right)e^{+i\frac{\omega}{2}t} + \left(\begin{array}{c}
\psi_{A}^{(1),-}\\
\psi_{B}^{(1),-}
\end{array}\right)e^{-i\frac{\omega}{2}t}\Biggr],
\label{res1-comparison}
\end{eqnarray}
where the upper sub-index $(1)$ refers to the first order resonance approximation. Following exactly the same approach as we used above, that is
substituting (\ref{res1-comparison}) into eq. (\ref{eq1}) and equating the amplitudes that multiply $e^{-i \omega t/2}$ and $e^{i \omega t/2}$ separately, while ignoring higher harmonics, we arrive at a simple linear matrix equation 
\begin{equation}
\varepsilon^{(1)} \bar{\psi}^{(1)}=L^{(1)}\bar{\psi}^{(1)} \label{mat-eq}
\end{equation} 
with a $4\times 4$ matrix $L^{(1)}$ and a time-independent `vector' 
$\bar{\psi}^{(1)}=(\psi_{A}^{1,+}, \psi_{A}^{(1),-}, \psi_{B}^{(1),+}, \psi_{B}^{(1),-})$. Solving this simple eigenvalue-eigenvector problem results in the spectrum $\varepsilon=\varepsilon^{(1)}(k_x,k_y)$ which has two Dirac points (see Fig. 6c, red points) at $k=\sqrt{k_{x}^{2}+k_{y}^{2}}\approx \omega/2 $ in the first order approximation. 

In the second order approximation, in addition to the harmonics that correspond to $\pm \omega/2$ we keep the $\pm 3\omega/2$ harmonics, thus, searching for $\psi_0$ in the form 
\begin{eqnarray}
\nonumber
\psi^{(2)}_{0} = e^{-i \varepsilon t}\Biggl[\left(\begin{array}{c}
\psi_{A}^{(2),++}\\
\psi_{B}^{(2),++}
\end{array}\right)e^{+i\frac{3\omega}{2}t} + \left(\begin{array}{c}
\psi_{A}^{(2),+-}\\
\psi_{B}^{(2),+-}
\end{array}\right)e^{+i\frac{\omega}{2}t} + \noindent \\ + \left(\begin{array}{c}
\psi_{A}^{(2),-+}\\
\psi_{B}^{(2),-+}
\end{array}\right)e^{-i\frac{\omega}{2}t} + \left(\begin{array}{c}
\psi_{A}^{(2)--}\\
\psi_{B}^{(2),--}
\end{array}\right)e^{-i\frac{3\omega}{2}t}\Biggr],\nonumber\\
\end{eqnarray}
with sub-index $(2)$ referring into the second order resonance approximation. This problem reduces to the eigenvalue-eigenvector problem 
\begin{equation}
\varepsilon^{(2)} \bar{\psi}^{(2)}=L^{(2)}\bar{\psi}^{(2)} \label{mat-eq1}
\end{equation} 
with an $8\times 8$ matrix $L^{(2)}$ and a vector $\bar{\psi}^{(2)}=(\psi_{A}^{(2),++}, \psi_{A}^{(2),+-}, \psi_{A}^{(2),-+}, \psi_{A}^{(2),--}, \psi_{B}^{(2),++}, \psi_{B}^{(2),+-}, \\ \psi_{B}^{(2),-+}, \psi_{B}^{(2),--})$. The spectrum $\varepsilon=\varepsilon^{(2)}(k_x,k_y)$ is shown in Fig. 6d which has two Dirac points at  $k=\sqrt{k_{x}^{2}+k_{y}^{2}}\approx \omega/2 $ and six more Dirac points at  $k=\sqrt{k_{x}^{2}+k_{y}^{2}}\approx 3\omega/2$. These Dirac points are shown in blue on Fig. 6c. The two Dirac points at $k=\sqrt{k_{x}^{2}+k_{y}^{2}}\approx \omega/2$ almost coincide in both the first and the second order
resonance approximation, indicating that two spectra are almost the same for $k\lesssim \omega/2$. Moreover, the higher order resonance approximation
allows a better calculation of the spectrum at higher momentum resulting in the opening of a gap and the appearance 
of six more Dirac points at $k=\sqrt{k_{x}^{2}+k_{y}^{2}}\approx 3\omega/2$. 

\begin{figure}[htb]
\includegraphics[width=9cm]{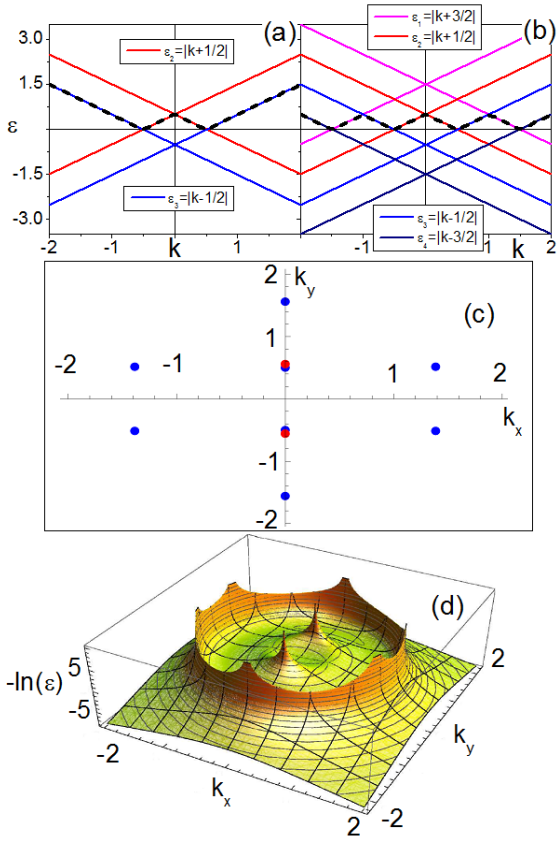}
\caption{(color online) Schematic representation of the photon shifted Dirac cones for the case of the first (a) and the second (b) order resonance approximation for monochromatic
laser field ${\bf A}=(0, A_1\cos\omega t)$. The black dots represent the lowest energy zone. (c) The Dirac points calculated in the first (red point)
and the second (blue points) order resonance approximation for $\omega=1, A_1=0.5$. The points for $k\approx \omega/2$ for the first and second order resonance approximation almost coincide. (d) 3D plot of the logarithm of the lowest energy zone: $-\ln\varepsilon(k_x,k_y)$ for the same parameters as in Fig. 6(c).}
\label{currents-super}
\end{figure}

The physical meaning of the different spectra obtained in the first and second resonance approximation can be interpreted in the limit of a very low amplitude of 
laser field. In the first order approximation, instead of the usual Dirac energy cone, we should consider the energy spectrum of a hole and a photon and
the spectrum of an electron and an emitted photon. This produces two shifted cones $\varepsilon\pm\omega/2$ (see red and blue energy spectra in fig 6a).
As a result, the new low energy zone  (black dots in fig. 6(a)) with zero-energy rings at $k=\omega/2$ forms in the limit $A_1\rightarrow 0$. Note that for 
$A_1=0$ this new zone structure is just an alternative way to represent an initial Dirac cone. This situation is similar to the extended and 
reduced zone representations for infinitesimally weak, spatially periodic potential. At finite $A_1$ two Dirac points of zero energy form instead of the zero energy rings (Fig. 6(c), red points). 
In the second order approximation we need to consider the initial Dirac cone shifted either by one or by two photons resulting in four shifted cones
(Fig. 6(b)) $\varepsilon\pm\omega/2$ and $\varepsilon\pm 3\omega/2$ producing two zero energy rings at $k=\omega/2$ and $k=3\omega/2$. 
Again the final field intensity $A_1\ne 0$ creates gaps in this energy spectrum and six more Dirac points at $k\approx 3\omega/2$. Obviously, calculation of
energy spectrum in higher order approximations will results in formation of other Dirac points at higher momentum.     

From the above analysis, we conclude that the first order resonance approximation for a monochromatic laser field ${\bf A}=(0, A_1\cos\omega t)$ 
can well describe the Dirac points and spectra for not too high Dirac fermion momentum $k\lesssim \omega/2$. 
Similar analysis allows the verification of the applicability of the resonance approximation (\ref{mu-firstorder}) for a bi-harmonic laser field (\ref{bi-field})
at $k\lesssim 3\omega/2$. This justifies the use of the method to calculate the spectra for low enough momentum $k$ as well as the evolution of the Dirac points with varying laser field parameters. 

\subsection{Spatial second order resonance approximation for superlattice}

We now consider the spatial second order resonance approximation for graphene superlattice with electrostatic field (\ref{pot}) driven by a monochromatic laser field 
 ${\bf A}=( A_1\cos\theta \cos\omega t, A_1\sin\theta\cos\omega t)$, thus,  keeping the exponentials 
$\exp\{\pm i\omega t/2\pm i\mu x/2\}$ and $\exp\{\pm i\omega t/2\pm i3\mu x/2\}$ in the expansion of ${\tilde \psi}_0(t)$. We are looking for the following approximate solution 
\begin{eqnarray}
\psi_0 &=& e^{-i \varepsilon t+ik_x x}\times\nonumber \\
&\times&\Biggl[\left(\begin{array}{c}
\psi_{A}^{2,+++}\\
\psi_{B}^{2,+++}
\end{array}\right)e^{+i\frac{\omega}{2}t+i\frac{3\mu}{2}x} + \left(\begin{array}{c}
\psi_{A}^{2,+--}\\
\psi_{B}^{2,+--}
\end{array}\right)e^{+i\frac{\omega}{2}t-i\frac{3\mu}{2}x} + \noindent \nonumber\\ 
&+& \left(\begin{array}{c}
\psi_{A}^{2,++}\\
\psi_{B}^{2,++}
\end{array}\right)e^{+i\frac{\omega}{2}t+i\frac{\mu}{2}x} + \left(\begin{array}{c}
\psi_{A}^{2,+-}\\
\psi_{B}^{2,+-}
\end{array}\right)e^{+i\frac{\omega}{2}t-i\frac{\mu}{2}x} + \noindent \nonumber\\ 
&+& \left(\begin{array}{c}
\psi_{A}^{2,-+}\\
\psi_{B}^{2,-+}
\end{array}\right)e^{-i\frac{\omega}{2}t+i\frac{\mu}{2}x} + \left(\begin{array}{c}
\psi_{A}^{2,--}\\
\psi_{B}^{2,--}
\end{array}\right)e^{-i\frac{\omega}{2}t-i\frac{\mu}{2}x}
+ \noindent \nonumber\\ 
&+& \left(\begin{array}{c}
\psi_{A}^{2,-++}\\
\psi_{B}^{2,-++}
\end{array}\right)e^{-i\frac{\omega}{2}t+i\frac{3\mu}{2}x} + \left(\begin{array}{c}
\psi_{A}^{2,---}\\
\psi_{B}^{2,---}
\end{array}\right)e^{-i\frac{\omega}{2}t-i\frac{3\mu}{2}x}
\Biggr]. \nonumber \\ \label{mu-second}
\end{eqnarray}
Substituting this trial function into the Dirac equation (\ref{dirac-super}) and ignoring all higher harmonics we reduce our system to the eigenvalue-eigenvector 
problem for 
\begin{equation}
\varepsilon^{(2s)} \bar{\psi}^{(2s)}=L^{(2s)}\bar{\psi}^{(2s)} \label{mat-eq2s}
\end{equation} 
where upper index $2s$ refers on the spatial second order resonance approximation described above, $L^{(2s)}$ is the corresponding 16$\times$16 matrix for
the following $x$-$t$-independent ``vector'' $\bar{\psi}^{(2s)}=(\psi_{A}^{2,+++},\psi_{B}^{2,+++}, \psi_{A}^{2,+--}, \psi_{B}^{2,+--}, \psi_{A}^{2,++}, \psi_{B}^{2,++},
\psi_{A}^{2,+-}, \newline\psi_{B}^{2,+-}, \psi_{A}^{2,-+}, \psi_{B}^{2,-+}, \psi_{A}^{2,--}, \psi_{B}^{2,--}, \psi_{A}^{2,-++}, \psi_{B}^{2,-++},  \newline\psi_{A}^{2,---},
\psi_{B}^{2,---})$. When increasing the rank of the matrix up to 16$\times$16, our numerical calculations become much more time consuming but the spectra $\varepsilon^{(2s)}$ can be calculated
and compared with the first resonance approximation described in  section \ref{sec-3}.

\begin{figure}[htb]
\includegraphics[width=9cm]{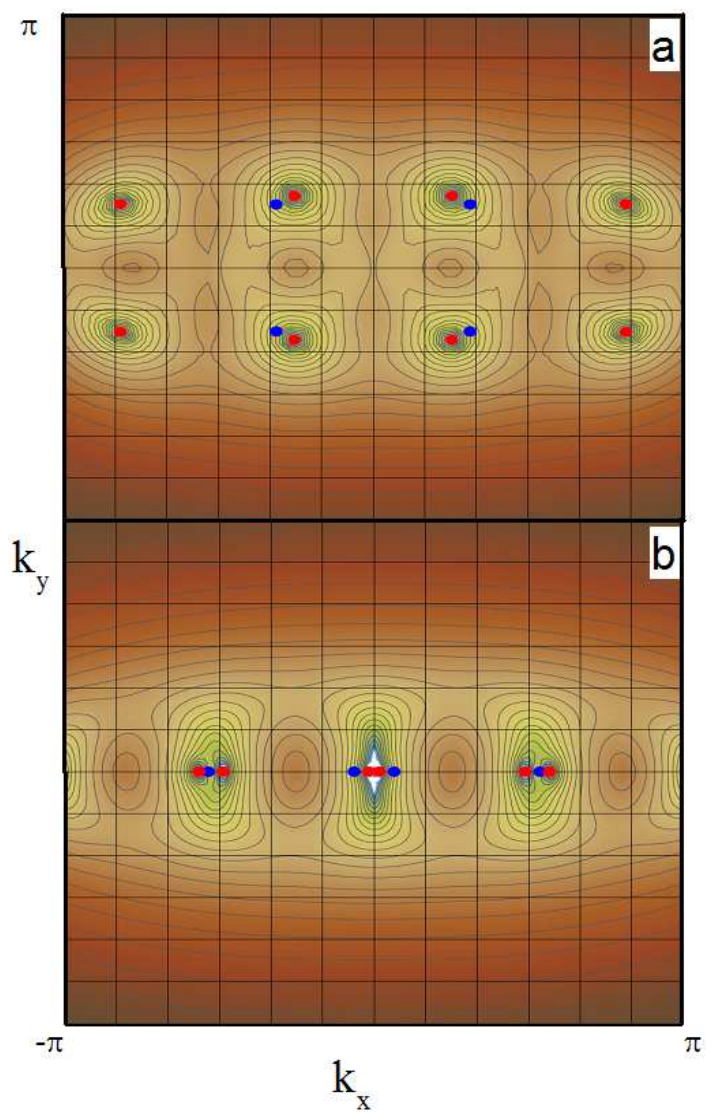}
\caption{(color online) (a,b) Contour plot of the lowest energy-band for a graphene superlattice with spatially periodic potential (\ref{pot}) while driven by the monochromatic laser field 
${\bf A}=( A_1\cos\theta \cos\omega t, A_1\sin\theta\cos\omega t)$, calculated using the spacial second order resonance approximation (\ref{mu-second}) for the same parameters as 
the energy spectrum shown in Fig. 4 for $\theta=\pi/2$ (a) and $\theta=0$ (b). The blue and red small circles show locations of Dirac points obtained within the spatial first order and the spatial 
second order resonance approximations.}
\label{currents-super}
\end{figure}

Comparing the energy spectra obtained in the first and second spatial resonance approximation (Fig. 7), we conclude that the higher order harmonics have very little effect when the electrostatic field $\nabla U$
and laser field $\propto d{\bf A}/dt$ are orthogonal (Fig. 7a). In this case the four low-momentum Dirac points are just slightly shifted from their positions when the higher order resonance approximation is used.
It is clear that new (four) Dirac points for higher momentum occur in the second order resonance approximation. For the case when both electrostatic and laser fields are directed along the same ($x-$) axis,
the higher harmonics influence the spectrum strongly: in addition to a simple shift of the two very-low momentum Dirac point, we observe splitting of the two higher momentum Dirac points (obtained in the first order approximation)
into four Dirac points (in the second approximation). Note that the split Dirac points are located at momentum just slightly below $3\mu/2$, where we can expect the stronger influence of higher harmonics. 

\section{Conclusion}

To conclude, in the present study we consider the quasi-energy spectra for both graphene and topological insulators and demonstrate that an application of the biharmonic laser field can provide a very useful method to manipulate the number and position of Dirac points in the spectrum where the two lowest energy zones merge. 

There are, certainly, many other ways to manipulates the number and position of the Dirac points in both classes of materials. The reason we make this specific proposal is that it is easy to be implemented experimentally and there is a good control, within the calculational scheme we use, of the results. It is evident that there is a strong effect on the single-particle current. We have shown how it manifests itself in the DC electrical current which, in turn, can be manipulated by the laser fields. For the case of graphene/TI superlattices, the Dirac points and single-particle current distributions can be well controlled even by a one-frequency (monochromatic) laser field 
if its electric field is rotated with respect to the gradient of the superlattice potential. However the DC electric current is expected to be zero due to the high symmetry of $j_{x,y}(k_x,k_y)$. This is a prediction that can be tested experimentally, even without invoking details beyond the single-particle picture and a full, more complicated calculation of the current.

\acknowledgments
This work has been supported by the Engineering and Physical Sciences Research Council under the grant EP/H049797/1, the Leverhulme Trust and the project MOSAICO.

\end{document}